\documentclass[preprint]{aastex}
\usepackage[latin1]{inputenc}
\usepackage{amssymb}
\usepackage{amsmath}
\usepackage{graphicx}
\usepackage{xspace}
\usepackage{natbib}
\usepackage{url}
\usepackage{paralist}
\usepackage{epsfig}

\def\cm{{\rm\thinspace cm}}

\def\erg{{\rm\thinspace erg}}

\def\Msun{\hbox{$\rm\thinspace M_{\odot}$}}

\def\s{{\rm\thinspace s}}

\def\chisq{\hbox{$\chi^2$}}

\def\ergpcmsqps{\hbox{$\erg\cm^{-2}\s^{-1}\,$}}

\def\ergps{\hbox{$\erg\s^{-1}\,$}}

\slugcomment{ }

\shorttitle{An XMM-Newton View of the Radio Galaxy 3C 411}
\shortauthors{}

\begin{document}

\title{An XMM-Newton View of the Radio Galaxy 3C 411}

\author{Allison~Bostrom\altaffilmark{1}, Christopher~S.~Reynolds\altaffilmark{1,2}, Francesco~Tombesi\altaffilmark{1,3}}
\email{}
\altaffiltext{1}{Department of Astronomy, University of Maryland, College Park, MD 20742-2421, USA}
\altaffiltext{2}{Joint Space-Science Institute (JSI), College Park, MD 20742-2421, USA}
\altaffiltext{3}{NASA Goddard Spaceflight Center, Greenbelt, MD, USA}

\begin{abstract}
We present the first high signal-to-noise XMM-Newton observations of the broad-line radio galaxy 3C 411.  After fitting various spectral models, an absorbed double power-law continuum and a blurred relativistic disk reflection model (kdblur) are found to be equally plausible descriptions of the data.  While the softer power-law component ($\Gamma$=2.11) of the double power-law model is entirely consistent with that found in Seyfert galaxies (and hence likely originates from a disk corona), the additional power law component is very hard ($\Gamma$=1.05); amongst the AGN zoo, only flat-spectrum radio quasars have such hard spectra. Together with the flat radio-spectrum displayed by this source, we suggest that it should instead be classified as a FSRQ. This leads to potential discrepancies regarding the jet inclination angle, with the radio morphology suggesting a large jet inclination but the FSRQ classification suggesting small inclinations. The kdblur model predicts an inner disk radius of at most 20 r$_g$ and relativistic reflection.

\end{abstract}



\section{Introduction}\label{intro}

X-ray spectroscopy is an important tool for understanding the origin of jets in active galactic nuclei (AGN).  Broad-line radio galaxies (BLRGs) represent a particularly important class of object when considering the radio-loud/radio-quiet (RL/RQ) dichotomy, as their properties resemble those of the well-studied Seyfert galaxies, but they are also jetted and radio-loud.  Though the X-ray spectral characteristics of BLRGs are generally not well known or understood, those that have been observed generally exhibit weaker Compton reflection features than Seyferts, and weaker iron K$\alpha$ emission lines (e.g., \citealt{wozniak:98}, \citealt{sambruna:99}, \citealt{reynolds:03}).  Several mechanisms have been proposed to explain this difference; one explanation is that the inner accretion disk is in a hot, advective state incapable of producing reflection features (\citealt{rees:82}, \citealt{narayan:98}), while another is that an unresolved jet component contributes to the X-ray spectrum of BLRGs, diluting the reflection features from the disk. \citet{kataoka:11} investigated the jet dilution explanation by fitting blazar-like models to the broad-band (radio to $\gamma$-ray) spectral energy distributions (SEDs) of several radio-loud AGN.  Among the BLRGs studied, the jet was generally found to constitute a non-negligible part of the total observed luminosity, perhaps supporting the dilution picture.  Another potential difference between BLRGs and their radio-quiet cousins is a higher level of ionization in BLRG accretion disks \citep{ballantyne:02}.  Understanding the subtle differences that place BLRGs and Seyfert galaxies on opposite sides of the RL/RQ divide may prove to be an important key in explaining the dichotomy in a more general sense.

Another aspect of the comparison between RL/RQ is represented by AGN winds. In particular, evidences for highly ionized, ultra-fast outflows (UFOs) with velocities $\sim0.1$c have been reported for several Seyfert galaxies (e.g., \citealt{tombesi:10a}, \citealt{tombesi:11a}; \citealt{gofford:13}). Recently, UFOs and the lower ionized/slower warm absorbers (WAs) have been reported in a few BLRGs as well (\citealt{reeves:09}, \citealt{tombesi:10b}, \citealt{tombesi:11b}, \citealt{tombesi:13a}, \citealt{torresi:10}, \citealt{torresi:12}, \citealt{gofford:13}). Importantly, the mechanical power from these winds seems to be sufficient to significantly contribute to the feedback of the AGN on the surrounding environment, adding to any contribution from the jets (e.g., \citealt{tombesi:12a}, \citealt{tombesi:13b}, \citealt{crenshaw:12}). Therefore, the search for wind absorption features in the spectra of radio galaxies, such as 3C~411, is fundamental to quantify their covering fraction and mechanical power and perform a comparison with jets and the winds in RQ sources (e.g., \citealt{tombesi:12b}).

The differences between Seyferts and BLRGs underscore the complexity involved in classifying AGN and illustrate that, though elegant, the unified model is incomplete, particularly in the radio-loud realm.  The unification of type 1 and type 2 Seyfert galaxies has been explained by postulating the presence of a torus of obscuring material around the nucleus, such that the optical broad-line region may be rendered invisible at higher viewing angles \citep{antonucci:85}.  It is not clear whether such a simple picture will hold for radio-loud AGN \citep{antonucci:12}.  For example, BL Lacs are believed to have lower line luminosities than flat-spectrum radio quasars (FSRQs), but line luminosities of blazars may well exist on a continuum rather than being divided into two separate groups \citep{urry:95}.  The same may be true for high-peak BL Lacs (HBLs) and low-peak BL Lacs (LBLs); instead of being two distinct classes, more recent data implies that perhaps there is a gradient of synchrotron peaks in BL Lac SEDs (\citealt{laurent:99}, \citealt{nieppola:06}).  There may be two separate classes of narrow-line RL AGN, one that possesses obscuring tori and one that does not, which would affect the presence or absence of a broad-line region, but this is as yet an unanswered question.

This paper discusses the analysis of the XMM-Newton spectrum of 3C 411, a Fanaroff-Riley II BLRG with a double-lobed radio structure and a jet connecting at least one of the lobes to the central hot spot \citep{spangler:84}.  Section~\ref{reduction} will discuss the data and the reduction process, while Section~\ref{analysis} will focus on the spectral analysis.  Section~\ref{discussion} will outline the main results of the analysis, along with a discussion of the physical implications of each model. In Section~\ref{conclusions}, we summarize our results.

\section{Observations and Reduction}\label{reduction}

XMM-Newton observed 3C 411 on November 17, 2011 for a total duration of 63 ks.  Here we present an analysis of the EPIC-pn and EPIC-MOS-1 and EPIC-MOS-2 data; the Reflection Grating Spectrometer (RGS) data had insufficient signal-to-noise to provide additional constraints.  The EPIC-pn and EPIC-MOS instruments were operated in full window imaging mode throughout the observation.  The reduction was performed using SAS version 1.52.9 according to the  procedure outlined in the XMM ABC Guide.  The standard filtering expressions were applied to the data, in addition to temporal filters to eliminate soft proton flaring contamination.  After filtering, the good on-source exposure time was 39.5 ks for pn and 60.7 ks for MOS.  The source spectrum was extracted in DS9 using a circular extraction region (of radius 38 arcseconds for the pn spectrum, 33 for the MOS-1 spectrum, and 37 for the MOS-2 spectrum), and the background spectrum was extracted from an annulus surrounding the source.  The response matrix file (rmf) and the ancillary response file (arf) were created using the SAS tools rmfgen and arfgen, respectively.  The spectra were grouped to a minimum of 20 photons per bin using the FTOOLS grppha tool in order to facilitate standard $\chi^2$ fitting techniques.

\section{Spectral Analysis}\label{analysis}

The spectral analysis was carried out with XSPEC version 12.7.1. The data from the three EPIC instruments were jointly analyzed, and counts from outside the energy range of 0.5 - 10 keV were ignored due to a poor signal-to-noise ratio and uncertain calibration at the edges of the instruments' energy range.  For the purposes of plotting, the data from each instrument were grouped with a minimum signal-to-noise ratio of 30 (subject to a maximum of 100 combined bins) so that structure in the residuals could be visually explored.   An absorbed power-law model was fit to the data, including a constant factor to normalize between the three detectors.  The value for the column density of neutral hydrogen (using the multiplicative phabs model component) was taken to be the Galactic value of $N_H=1.34 \times 10^{21} {\hbox{$\cm^{-2}\,$}}$ \citep{dickey:90}.  As can be seen in Figure 1, the ratios clearly illustrate the presence of spectral complexities beyond a simple absorbed power-law ($\chisq/dof \approx 1.05$ for 1734 degrees of freedom, $\Gamma=1.93$).

\begin{figure*}[h !]
\centerline{
\psfig{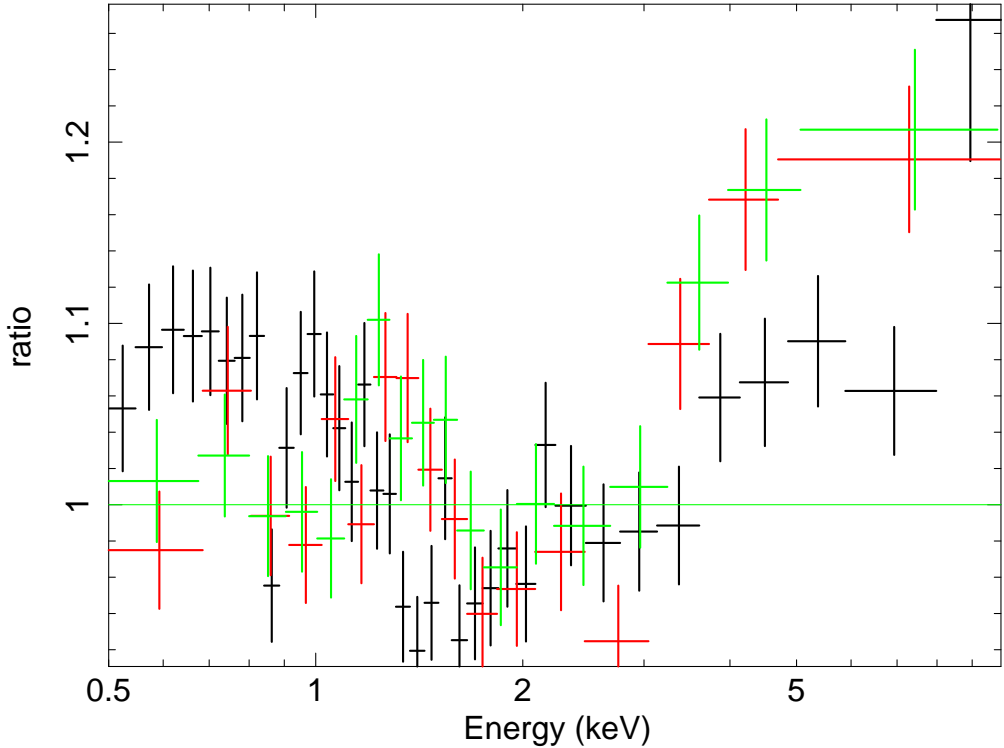}
}
\caption{The ratios of the absorbed single power-law model clearly indicate that the XMM-Newton X-ray spectrum of 3C 411 includes structure beyond this basic picture. The black points represent the data from the pn detector, the red from the MOS-1 detector, and the green from the MOS-2 detector.}
\label{fig:ratios}
\end{figure*}

To better understand this structure, several other models were fit to the data. An absorbed power-law/blackbody model was explored next.  This model includes a power-law component and its normalization added to a blackbody, all of which is multiplied by the same Galactic $N_H$ value through the phabs model component (see Figure 2 for a plot of the data and the model).  Though this model fits the data very well ($\chisq/dof \approx 0.99$ for 1732 degrees of freedom; $\Delta\chi^2=93$ for two additional degrees of freedom compared with the absorbed power-law model), it was discarded due to a lack of physical motivation.  The most plausible origin of a blackbody in an AGN is the accretion disk, but the disk usually peaks at $\sim$ 10 eV, which is much too cold for the 2 keV peak seen in our model. Even very hot disks usually only approach roughly 0.1 keV.  Thus we do not believe that the power-law and blackbody model, though a good fit, is the most accurate representation of our data.

\begin{figure*}[h !]
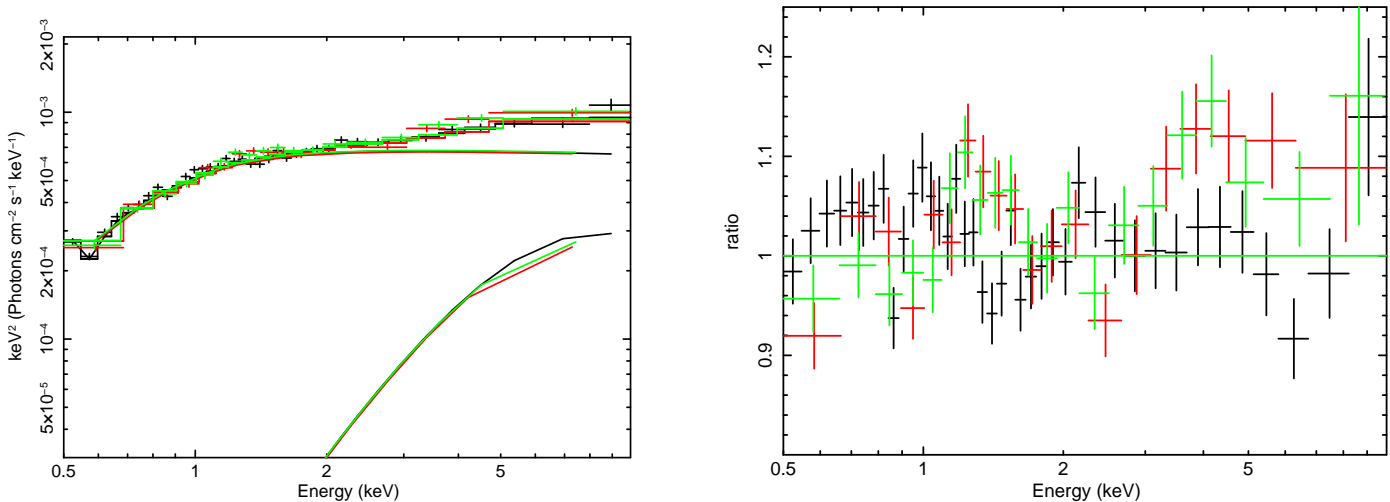

\centerline{
\hbox{
\psfig{figure=f2a.eps,width=0.4\textwidth,angle=270}
\hspace{0.5 cm}
\psfig{figure=f2b,width=0.4\textwidth,angle=270}
}}
\caption{ \emph{Left:} The data for all three detectors and the power law/blackbody model in solid lines. \emph{Right:} The data:model ratio for the power law/blackbody model, with unity represented by the horizontal green line. Both follow the same color scheme as Fig. 1.}
\label{fig:bbod}
\end{figure*}

An absorbed double power-law model was fit to the data next, and was found to describe the data very well ($\chisq/dof \approx 1$ for 1732 degrees of freedom; $\Delta\chi^2=83$ for two additional degrees of freedom compared with the absorbed power-law model).  This model consists of a soft Seyfert-like component and a much harder component more similar to those typically found in blazars, shown in Figure 3. A more in-depth discussion of the double power-law model can be found in the following section.

\begin{figure*}[h !]
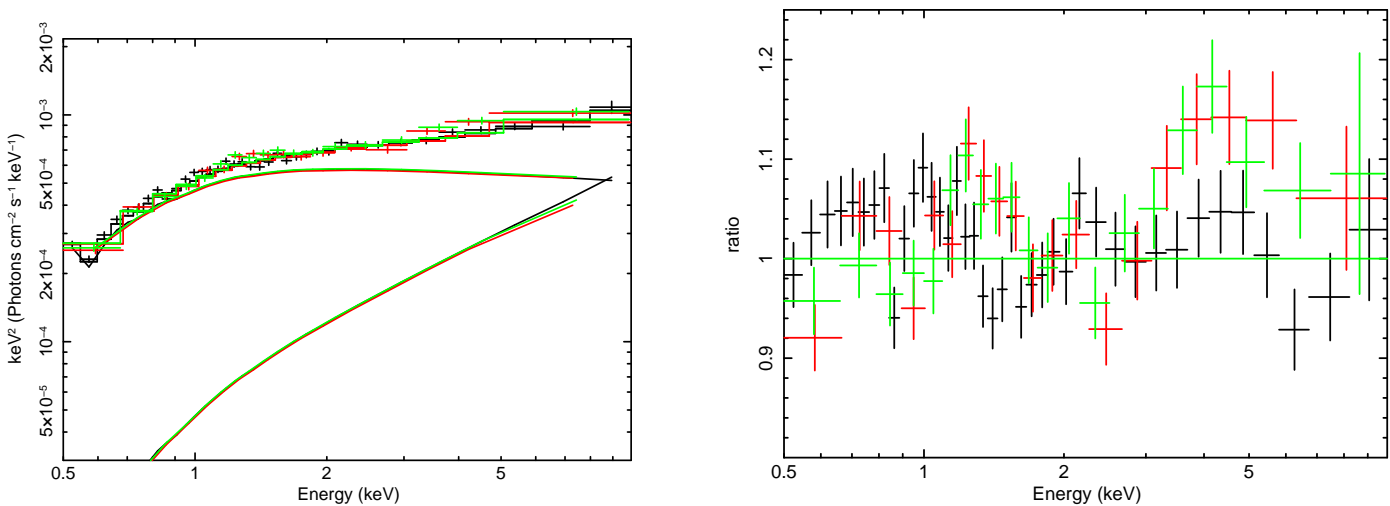

\centerline{
\hbox{
\psfig{figure=f3a.eps,width=0.4\textwidth,angle=270}
\hspace{0.5 cm}
\psfig{figure=f3b.eps,width=0.4\textwidth,angle=270}
}}
\caption{ \emph{Left:} The data (crosses) and the double power law model (solid lines). \emph{Right:} The data:model ratio for the double power law model, with unity represented by the horizontal green line. Both follow the same color scheme as Fig. 1.}
\label{fig:2pl}
\end{figure*}

The next model we fit to our data represents an irradiated accretion disk around a rotating black hole, creating a reflection spectrum which is then smeared due to relativistic effects. This was accomplished by first modeling the power-law reflection spectrum with the reflionx model, then convolving it with the kdblur model to account for the smearing due to gravitational redshift and high velocities surrounding a spinning black hole.  The kdblur model was adapted by Fabian and Johnstone from the Laor convolution model, and includes Galactic absorption, as well as an emissivity index (fixed at $q=3.00$), inner and outer disk radii (the latter was fixed at 100.00 r$_g$), and the disk inclination angle \citep{laor:91}.  The model also accounts for the iron abundance of the disk relative to the solar abundance (fixed at unity), ionization ($\xi$), and the source redshift (fixed at 0.467).  The power-law photon index and its normalization are also included.  This model also fits the data very well  ($\chisq/dof \approx 0.99$ for 1730 degrees of freedom; $\Delta\chi^2=98$ for 4 additional degrees of freedom).   A comparison of this model to the data is shown in Figure~4.  While this model has a better goodness of fit parameter than the double power law model, the fact that these are not nested models prevents a direct statistical comparison using, for example, the F-test.  In fact, both models provide a statistically valid description of the data.

\begin{figure*}[h !]
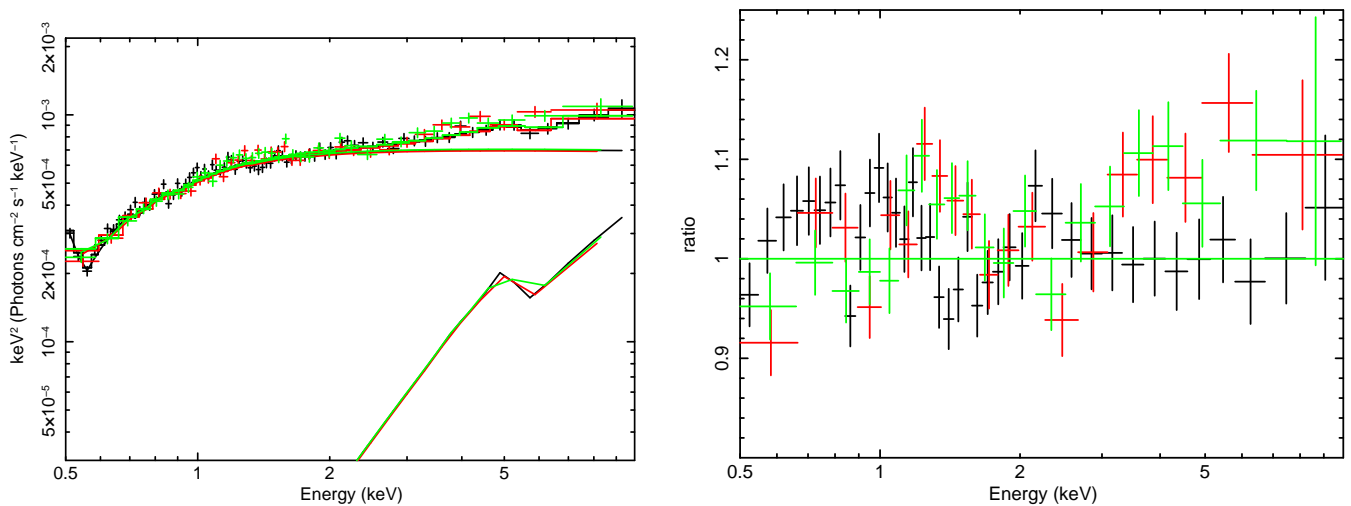

\centerline{
\hbox{
\psfig{figure=f4a.eps,width=0.4\textwidth,angle=270}
\psfig{figure=f4b.eps,width=0.4\textwidth,angle=270}
}}
\caption{ \emph{Left:} The data (crosses) and the kdblur model (solid lines). For this plot, the data were binned to a minimum signal-to-noise ratio of 20, with a maximum of 20 combined bins, to better visualize the shape of the model. \emph{Right:} The data:model ratio for the kdblur model, with unity represented by the horizontal green line. Both follow the same color scheme as Fig 1.}
\label{fig:kdblur}
\end{figure*}

Finally, we used {\sc Xstar} photoionization table models to test for the presence of warm absorbers.  We fit models for an absorber at systemic velocity, and one moving at 0.1c.  We added this component to our double power-law fit.  The {\sc Xstar} models include the column density of the absorber, an ionization parameter log($\xi$) which was frozen, and a redshift component. The redshift component can be modified to test for outflowing absorbers of different velocities.

In all models, the Galactic $N_H$ value was frozen.  Table 1 outlines the parameter values for each model.  We fit each of the models to the pn and MOS data separately to ensure that the discrepancies between the detectors did not significantly affect the parameter values, and found parameter values consistent with those obtained by our joint analysis. The flux of the source in the 2 - 10 keV band using the double power-law model is $2.20 \pm 0.16 \times 10^{-12} \ergpcmsqps$, which, at the luminosity distance of 2.52~Gpc (corresponding to the source redshift of $z=0.467$ in a cosmology with $H_0=73\,{\rm km}\,{\rm s}^{-1}\,{\rm Mpc}^{-1}$, $\Omega_M=0.27$ and $\Omega_\Lambda=0.73$), yields a luminosity of $\sim5 \times 10^{45} \ergps$ over a few decades in energy. This luminosity is approximately the Eddington luminosity of a $3 \times 10^7 \Msun$ black hole, giving a lower limit on the black hole mass. We derived an upper limit to the black hole mass from the V-band magnitude given in \citet{varano:04} and the black hole mass - bulge V-band absolute magnitude correlation found by \citet{wu:2001}. By assuming that all of the V-band luminosity originates in the bulge, we found an upper limit to the black hole mass of $\sim10^{10} \Msun$. The models are discussed further in Section 4.

\begin{table}[h !]
\begin{center}
    \begin{tabular}{llll}\hline\hline
    Model        & Parameter     & Value     & \chisq/dof    \\ \hline
    PL  & $\Gamma$      & $1.93\pm0.011$    &1812.12/1734       \\
    ~		& Normalization	& $7.25\times10^{-4}$	\\ \hline
    PL+BB      & $\Gamma$      & $2.04^{+0.028}_{-0.027}$   &1719.16/1732       \\
    ~		& Photon Index Normalization	& $7.24\times10^{-4}$	\\
    ~            & Temperature   & $2.09^{+0.399}_{-0.251}$ keV     \\
    ~		& Temperature Normalization	& $7.74\times10^{-6}$	\\ \hline
    2PL & Soft $\Gamma$ & $2.11^{+0.156}_{-0.077}$  &1729.01/1732        \\
    ~		& Soft Normalization		& $6.59\times10^{-4}$	\\
    ~            & Hard $\Gamma$ & $1.05^{+0.398}_{-0.497}$          \\
    ~		& Hard Normalization	& $6.74\times10^{-5}$	\\ \hline
    kdblur       & $\Gamma$      &  $2.01^{+0.0152}_{-0.0188}$    & 1714.19/1730     \\
    ~		& Normalization	& $7.3\times10^{-4}$    \\
    ~            & R$_{in}$          &  20 r$_g$ (upper limit)    \\
    ~            & Inclination   & $58^{+22}_{-33}$ $^\circ$ \\
    ~		& Fe/solar		& 1.0 (frozen)       \\
    ~		& $\xi$			&  1.24 (upper limit)	\\ \hline
    {\sc Xstar} 100 km/s &   Column Density   &   $6.69\times10^{20} {\hbox{$\cm^{-2}\,$}}$ (upper limit)   &  1729.35/1731	\\ 	
    ~		& log($\xi$) 		& 2.00 (frozen)	&  	\\	
    {\sc Xstar} 1000 km/s &  Column Density  &  $1.45\times10^{22} {\hbox{$\cm^{-2}\,$}}$ (upper limit)	  &  1729.02/1731 \\
    ~		& log($\xi$)		& 4.00 (frozen) & 	\\  \hline \hline
    \end{tabular}
    \end{center}
\caption{Parameter values for the various models (all of which were subjected to Galactic absorption $N_H=1.34 \times 10^{21} {\hbox{$\cm^{-2}\,$}}$): power-law (PL), double power-law (2PL), power-law/blackbody (PL+BB), accretion disk reflection  (kdblur), and {\sc Xstar} wind absorption.}
\end{table}

\section{Discussion}\label{discussion}

After finding two equally well-fitting models, we explored the physical implications of each.  The kdblur model implies a cold accretion disk with reflection from a hot corona and fairly standard parameter values for a BLRG. The double power-law model consists of a Seyfert-like component ($\Gamma=2.11$) and a very hard component ($\Gamma=1.05$).  Here, we hypothesize that the latter component might originate from the jet, and after delving further into this picture, find evidence for blazar-like behavior from 3C 411.

The double power-law model implies a curious structure for 3C 411. The steep power-law photon index is consistent with a typical Seyfert power-law ($\Gamma=2.11$).  This power-law component arises from inverse-Compton scattering of disk photons by electrons in the hot corona.  The other power-law photon index is very hard ($\Gamma=1.05$), and found to be comparable to those of FSRQs, in which the jet is aligned at a very small angle to the line of sight \citep{chen:13}.  The SED found on the NASA/IPAC Extragalactic Database (NED) corroborates this view, as the radio spectrum is flat, a feature that is characteristic of FSRQs and, indeed, is the origin of their name (see Figure 5).  3C 411 was determined to have a high "mixing parameter" relative to the other BLRGs in the sample of \citet{kataoka:11}, indicating a higher jet contribution to the overall luminosity, which supports the hard power-law index found in our analysis.  However, this blazar view of 3C 411 is very puzzling given its "classic double" radio morphology, which is indicative of a large jet inclination angle: it has a central radio core with well defined lobes on either side, connected by bridges of radio emission \citep{spinrad:75}.  Because FSRQs are typically believed to have small jet inclinations, these two pictures seem to be incompatible.  One of the lobes is brighter than the other, suggesting that perhaps the jet orientation could be somewhat tilted with respect to the plane of the sky, but it is unclear by how much a jet could be angled and still appear as a blazar. Therefore, whether it is possible that the galaxy is tilted just enough to still appear as a blazar but also be projected on the sky as a radio galaxy is unknown, although it seems unlikely.  Another possible explanation for the discrepancy is a reorienting jet, such that at one time the jet is oriented perpendicular to the line of sight, and then at a later time is pointed along the line of sight.  As such, the classification of 3C 411 poses an interesting challenge to the unified model, in which differences between AGN types are explained by variations in observer viewing angle.

\begin{figure*}[h !]
\centerline{
\psfig{figure=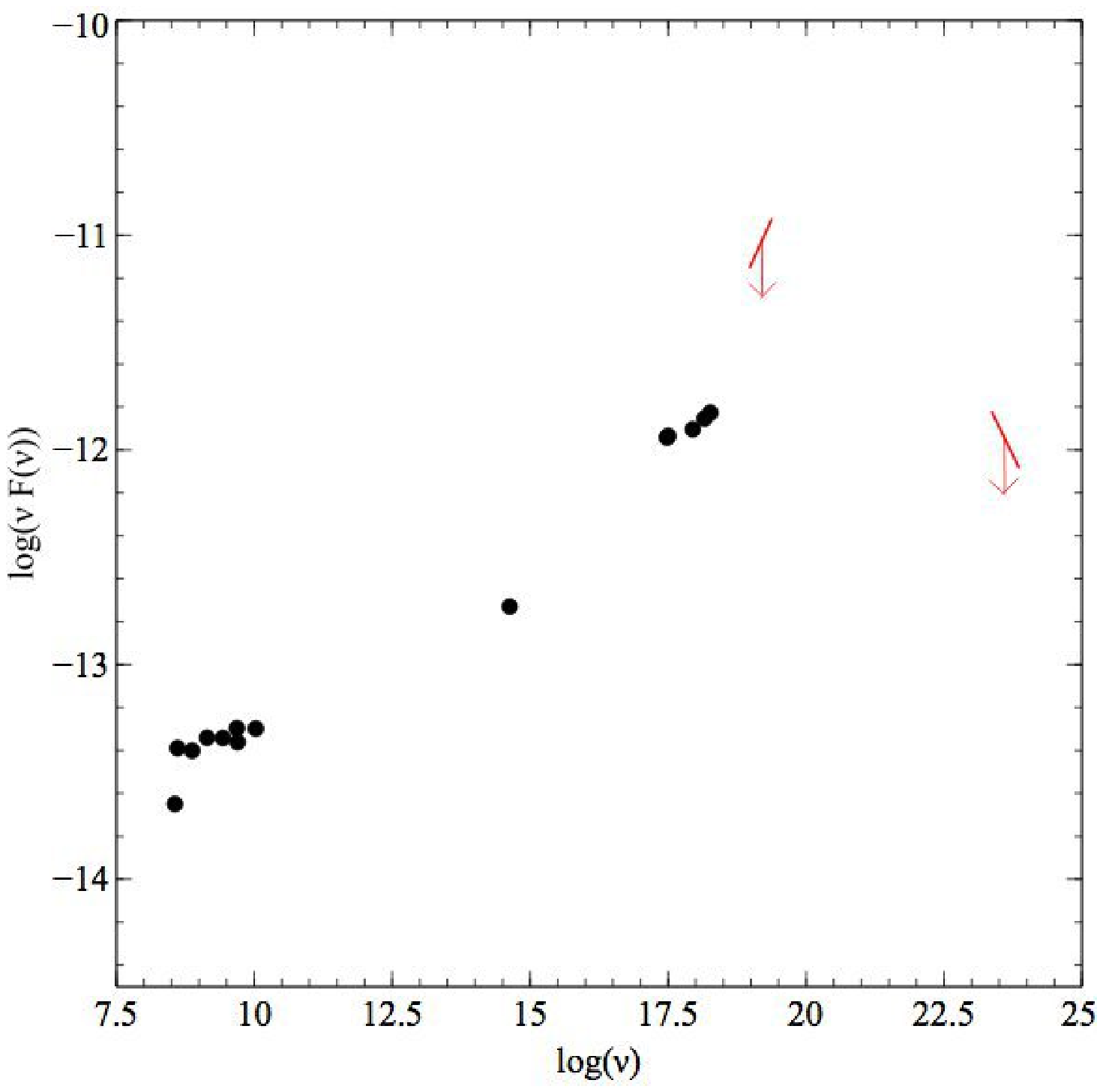,width=0.7\textwidth}
}
\caption{3C 411 exhibits a  flat radio spectrum, as is evident from its SED. This supports the blazar picture of 3C 411. The hard X-ray limit here was derived from the fact that 3C 411 was not detected by the Swift Burst Alert Telescope (BAT), which has a sensitivity of $10^{-8} \ergpcmsqps$. The gamma-ray upper limit was taken from \citet{kataoka:11}. The optical point was part of a study by \citet{varano:04}, and the radio points were taken from \citet{kellerman:73}, \citet{kellerman:69}, \citet{gregory:91}, \citet{pauliny-toth:65}, \citet{large:81}, \citet{douglas:96}, and \citet{gower:67}.  Fluxes are in $\ergpcmsqps$ and frequencies are in Hz.}
\label{fig:sed}
\end{figure*}

The accretion disk reflection (kdblur) model provides a description of the data that does not challenge 3C 411's classification. This model includes a Seyfert-like power-law component typical of BLRGs ($\Gamma=2.02$) and predicts an inclination angle between 25$^\circ$ and 80$^\circ$.  Though the inclination is not well-constrained, these constraints are consistent with the inclination angles expected to produce broad lines, in agreement with 3C 411's classification as a broad-line radio galaxy. The inner radius upper limit of 20 r$_g$ suggests that the spectrum is subject to relativistic reflection.

As discussed in the Introduction, it is interesting to search for warm absorbers in this source.  To test for the presence of warm absorbers, we constructed a tabulated (multiplicative) absorption model using the {\sc Xstar} code.  The absorbing gas is characterized by its ionization parameter $\xi$ and column density $N_H$.   Initially, the absorption was assumed to be at rest with respect to the source ($z=0.467$).   Shown in Figure~6 (thin blue arrows) are the upper limits on the derived column densities as we step through a set of $\log\xi$ from $\log\xi=0$ to 4.  We also investigate the limits when the absorption is assumed to be outflowing with $v=0.1c$.  The results are fairly stringent constraints on warm absorber column densities for low to medium ionization parameters, with most of the upper limits falling below $10^{21} {\hbox{$\cm^{-2}\,$}}$.

\begin{figure*}[h !]
\centerline{
\psfig{figure=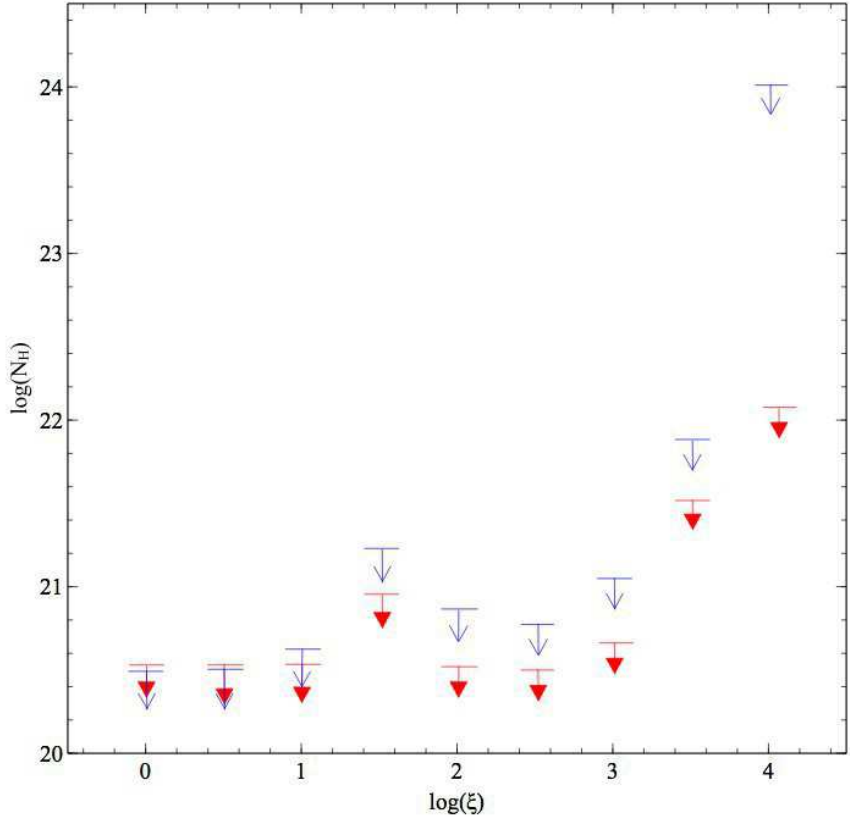,width=0.7\textwidth}
}
\caption{Column densities or upper limits for log($\xi$) from 0 to 4 in steps of log($\xi$)=0.5, for a warm absorber at systemic velocity and one outflowing at 0.1c. Thin blue arrows indicate upper limits for the absorber at systemic velocity, and red filled in arrows represent upper limits for the UFO. The column density for log($\xi$)=4.0 for the absorber at systemic velocity was completely unconstrained, and thus we had to put the upper limit at the upper boundary of the model, $10^{24} {\hbox{$\cm^{-2}\,$}}$.}
\label{fig:warmabs}
\end{figure*}

\section{Conclusions}\label{conclusions}

Both of our plausible X-ray spectral models for 3C~411 paint an interesting picture.  In the jet-dominated picture (double PL fit), we find a source that presents a significant challenge to the unification picture.  In the disk-dominated picture (PL+kdblur fit), we find evidence for relativistic reflection.   Unfortunately, given the current data available for 3C 411, we are unable to decide between these models. Future paths forward include a study of the AGN's optical polarization, which would help constrain the contribution of the jet component.   Sensitive hard X-ray spectroscopy, such as is possible with {\it NuSTAR} could also provide a powerful way to distinguish these models.  In addition, better X-ray data may allow us to further constrain the inner disk radius, which could help determine whether the SMBH in 3C 411 is spinning. If so, we could constrain the spin parameter, making 3C~411 only the second BLRG to have a well-determined iron-line based spin constraint \citep{reynolds:13a}. . In the disk-dominated picture, we would see a Compton reflection hump, peaking at 20--30\,keV and then falling again towards the end of the {\it NuSTAR} band, 80\,keV.   In the jet dominated picture, the hard X-ray power law, corresponding to inverse Compton emission from the jet  \citep{urry:99}, would be expected to continue unbroken through the {\it NuSTAR} band (up to 80\,keV).  Such a component would peak in the MeV range, making 3C 411 one of a small number of known MeV blazars.  This is consistent with a prediction made by \citet{kataoka:11}, who claimed that BLRGs should exhibit non-thermal MeV emission, which could comprise a few percent of their hard X-ray power.  The SED in Fig.~5 motivates such an MeV peak with the upward-slanting X-ray slope and lower gamma-ray upper limit.  Understanding this class of object has important implications. Currently it is unknown whether blazars exhibit a bimodal distribution in the form of HBLs and LBLs, since so few have been found to peak in the MeV range.  Increasing the number of known MeV blazars would shift the view from a bimodal structure to that of a continuum, where blazars can peak anywhere along a range of the spectrum.  FSRQs in general, the class of blazar 3C 411 most closely resembles, are important for their role as a possible source of the cosmic X-ray background (CXB), of which they could comprise a substantial portion.  FSRQs have been hypothesized to comprise the majority of the CXB above 500 keV \citep{ajello:09}. Thus, a more conclusive classification for 3C 411 may prove beneficial for numerous fields of study.



\acknowledgments
\section*{Acknowledgments}

We thank David Ballantyne for interesting and fruitful discussions.  CSR thanks NASA for support under the ADAP program (grant NNX12AE13G).

\bibliographystyle{jwapjbib}
\bibliography{allison}

\end{document}